\documentclass[a4paper]{emulateapj}
\usepackage{graphicx}
\usepackage{amssymb}
\usepackage{color}
\usepackage{threeparttable}

\newcommand{\simgt}{\lower.5ex\hbox{$\; \buildrel > \over \sim \;$}}
\newcommand{\simlt}{\lower.5ex\hbox{$\; \buildrel < \over \sim \;$}}

\def\hkpc{\mathrel{h^{-1}{\rm kpc}}}
\def\hMpc{\mathrel{h^{-1}{\rm Mpc}}}
\def\Mpc{\mathrel{\rm Mpc}}
\def\Mvir{\mathrel{M_{\rm vir}}}
\def\cvir{\mathrel{c_{\rm vir}}}
\def\rvir{\mathrel{r_{\rm vir}}}
\def\Dvir{\mathrel{\Delta_{\rm vir}}}

\def\kms{\mathrel{\rm km\,s^{-1}}}
\def\hMsol{\mathrel{h^{-1}M_\odot}}

\defcitealias{Okabeetal:10}{Ok10}

\begin{document}

\title{LoCuSS: The Mass Density Profile of Massive Galaxy Clusters at z=0.2}

\author{
  Nobuhiro Okabe\altaffilmark{1},
  Graham P.\ Smith\altaffilmark{2},
  Keiichi Umetsu\altaffilmark{1}, 
  Masahiro Takada\altaffilmark{3}, and
  Toshifumi Futamase\altaffilmark{4}
}

\altaffiltext{1}{Academia Sinica Institute of Astronomy and
  Astrophysics (ASIAA), P.\ O.\ Box 23-141, Taipei 10617, Taiwan;
  okabe@asiaa.sinica.edu.tw}

\altaffiltext{2}{School of Physics and Astronomy, University of
  Birmingham, Edgbaston, Birmingham, B15~2TT, UK;
  gps@star.sr.bham.ac.uk}

\altaffiltext{3}{Kavli Institute for the Physics and Mathematics of
  the Universe (Kavli IPMU, WPI), The University of Tokyo, Chiba
  277-8582, Japan}

\altaffiltext{4}{Astronomical Institute, Tohoku University, Aramaki,
  Aoba-ku, Sendai, 980-8578, Japan}

\begin{abstract}
  We present a stacked weak-lensing analysis of an approximately
  mass-selected sample of 50 galaxy clusters at $0.15<z<0.3$, based on
  observations with Suprime-Cam on the Subaru Telescope\footnote{Based
    in part on observations obtained at the Subaru Observatory under
    the Time Exchange program operated between the Gemini Observatory
    and the Subaru Observatory.}
  \footnote{ Based in part on data collected at Subaru Telescope and
    obtained from the SMOKA, which is operated by the Astronomy Data
    Center, National Astronomical Observatory of Japan.}.
  We develop a new method for selecting lensed background galaxies
  from which we estimate that our sample of red background galaxies
  suffers just $1\%$ contamination.  We detect the stacked tangential
  shear signal from the full sample of 50 clusters, based on this
    red sample of background galaxies, at a total
  signal-to-noise ratio of ${\rm S/N}=32.7$.  The Navarro-Frenk-White
  model is an excellent fit to the data, yielding sub-10\% statistical
  precision on mass and concentration:
  $\Mvir=7.19^{+0.53}_{-0.50}\times10^{14}\hMsol$,
  $\cvir=5.41^{+0.49}_{-0.45}$ ($c_{200}=4.22^{+0.40}_{-0.36}$).
  Tests of a range of possible systematic errors, including shear
  calibration and stacking-related issues, indicate that they are
  sub-dominant to the statistical errors.  The concentration
    parameter obtained from stacking our approximately mass-selected
    cluster sample is broadly in line with theoretical predictions.
    Moreover, the uncertainty on our measurement is comparable with
    the differences between the different predictions in the
    literature.  Overall our results highlight the potential for
    stacked weak-lensing methods to probe the mean mass density
    profile of cluster-scale dark matter halos with upcoming surveys,
    including Hyper-Suprime-Cam, Dark Energy Survey, and KIDS.
\end{abstract}

\keywords{galaxies: clusters: general --- gravitational lensing: weak}


\section{Introduction}

  Gravitational lensing is a powerful probe of the matter
  distribution in galaxy clusters, because the observed signal is
  sensitive to the total matter distribution and insensitive to the
  physical processes at play within clusters.  Many studies have
  therefore employed gravitational lensing to probe the mass and
  internal structure of galaxy clusters \citep[][and references
  therein]{2011A&ARv..19...47K}.  Prominent among these studies are
  those that aim to measure the dependence of cluster density on
  cluster-centric radius, i.e.\ the ``density profile'' of clusters
  \citep[e.g.][]{1995ApJ...438..514M, 2001ApJ...552..493S,
    Gavazzietal:03, Kneibetal:03, 2003ApJ...588L..73D,
    2004ApJ...604...88S, Broadhurstetal:05, 2007ApJ...668..643L,
    Johnstonetal:07, Ogurietal:09, Okabeetal:10, Umetsuetal2011,
    Oguri12, 2013ApJ...765...24N}.  A major motivation is to test key
  predictions from the cold dark matter theory of structure formation:
  (1) the density profile of the dark matter halos posited to host
  galaxies and cluster of galaxies is predicted to be universal and
  follow a simple 2-parameter model \citep{NFW97}, and (2) massive
  galaxy cluster-scale dark matter halos have
  concentrations\footnote{$r_{200}$ is the radius within which the
    mean density is $200\times$ the critical density of the universe,
    and $r_{\rm s}$ is a ``scale radius'' at which $d\log\rho/d\log
    r=-2$.} of $c_{200}\equiv r_{200}/r_{\rm s}\simeq3-4$
  \citep[e.g.][]{Bullocketal:01, Dolagetal:04, Netoetal:07,
    Duffyetal:08, Zhaoetal2009}, and are thus ``less concentrated''
  than less massive halos.

  In order to probe the density profile across a large dynamic
  range, lensing studies have typically combined weak- and
  strong-lensing signals, and have thus been limited to small samples
  of strong-lensing-selected clusters.  These studies typically find
  that strong-lensing clusters have high central concentrations in
  projection \citep[e.g.][]{Gavazzietal:03, Kneibetal:03,
    Broadhurstetal:08, Umetsuetal2011, Oguri12}.  Moreover, joint
  lensing and dynamical studies find that the density profile of the
  dark matter component may be shallower than predicted from cold dark
  matter simulations \citep{2013ApJ...765...24N}.  Interpretation of
  these apparent tensions between observations and theory is
  complicated by possible selection biases, small sample size,
  lensing-projection bias caused by halo triaxiality, and the absence
  of baryons from the simulations upon which the predictions are
  based.

  We adopt a complementary approach that aims to make progress on
  overcoming issues relating to sample size and selection, and lensing
  projection biases.  Building on earlier stacked lensing studies
  \citep[][hereafter Ok10,]{2003ApJ...588L..73D, Johnstonetal:07, Okabeetal:10,
    Umetsuetal2011, Oguri12}, we measure the mean density profile of
massive clusters by stacking the weak-lensing signal from a sample of
50 approximately mass-selected clusters.  Our sample comprises
\emph{all} clusters from the \emph{ROSAT} All Sky Survey catalogs
\citep{Ebelingetal:98, Ebelingetal:2000, Bohringeretal:04} that
satisfy $L_X[0.1-2.4{\rm keV}]/E(z)^{2.7}\ge 4.2\times10^{44}\,{\rm
  erg\,s^{-1}}$, $0.15\le z\le 0.30$, $n_H<7\times10^{20}{\rm
  cm^{-2}}$, and $-25^\circ<\delta<+65^\circ$, where $E(z)\equiv
H(z)/H_0$ is the normalized Hubble expansion rate, and selecting on
$L_X/E(z)^{2.7}$ mimics a mass selection \citep{Popessoetal:2005}.  We
stress that our results are based on the \emph{full sample} of 50
clusters; sub-samples of clusters will be discussed in future
  articles.

In Section \ref{sec:analysis} we describe our data and analysis; in
Section \ref{sec:results} we explain our results and compare with numerical
simulations;and in Section \ref{sec:summary} we summarize our conclusions.  We
use the concordance $\Lambda$CDM model of $\Omega_{\rm M,0}=0.27$,
$\Omega_{\Lambda}=0.73$ and $H_0=100h\,\kms\,\Mpc^{-1}$ \citep{WMAP7}.
In this cosmology the virial over-density at the mean redshift of our
cluster sample, $\langle z\rangle=0.23$ is $\Delta_{\rm vir}=113.77$.
All error bars are 68\% confidence intervals unless otherwise stated.

\section{Subaru Data and Weak-lensing Analysis} \label{sec:analysis}

We observed all 50 clusters with Suprime-Cam \citep{Miyazakietal:02}
on the Subaru Telescope, as part of the Local Cluster Substructure
Survey (LoCuSS\footnote{\url{http://www.sr.bham.ac.uk/locuss}}) -- 46
clusters through the $V/i'$-band filters; two each through the $V/I_{\rm
  C}$- and $g/i'$-band filters.  Hereafter we refer to the bluer
filter as $V$ and the redder filter as $i'$.  The full-width half
maximum of point sources in the $V/i'$ bands is $0\farcs6\simlt{\rm
  FWHM}\simlt0\farcs9$ and $0\farcs5\simlt{\rm FWHM}\simlt0\farcs7$,
respectively.  Photometric calibration to $\le10\%$ precision in both
filters was achieved via observations of Landolt standard stars, and
double checked against SDSS/DR8 stellar photometry \citep{SDSSDR8}.

We measure the shape of faint galaxies using a modified version of
\citetalias{Okabeetal:10}'s pipeline, based on the {\sc
  imcat}\footnote{http://www.ifa.hawaii/kaiser/IMCAT}
\citep[][hereafter, KSB]{KSB}.  The main modification is to calibrate
the KSB isotropic correction factor for individual objects using
galaxies detected with high significance $\nu>30$ \citep{Umetsu+2010}.
This minimizes the inherent shear calibration bias in KSB$+$ methods
in the presence of measurement errors \citep{Okura+Futamase2012}.

We define a sample of background galaxies based on color with respect
to the red sequence of early-type galaxies in each cluster.  In
principle selecting red galaxies ($\Delta C\equiv(V-i')-(V-i')_{\rm
  ES0}>0$) yields a clean sample of background galaxies.  In reality a
positive color cut is required to eliminate contamination by faint red
cluster galaxies due to statistical errors and possible intrinsic
scatter in galaxy colors \citep[e.g.][]{Broadhurstetal:05}.  In
contrast, interpretation of ``blue galaxy'' ($\Delta C<0$) samples is
complicated by star-formation.  For completeness, we include blue
  galaxies in this section, however our results in Section \ref{sec:results}
  are based only on the red galaxy sample.

\begin{figure*}
  \centerline{
    \includegraphics[width=1\textwidth]{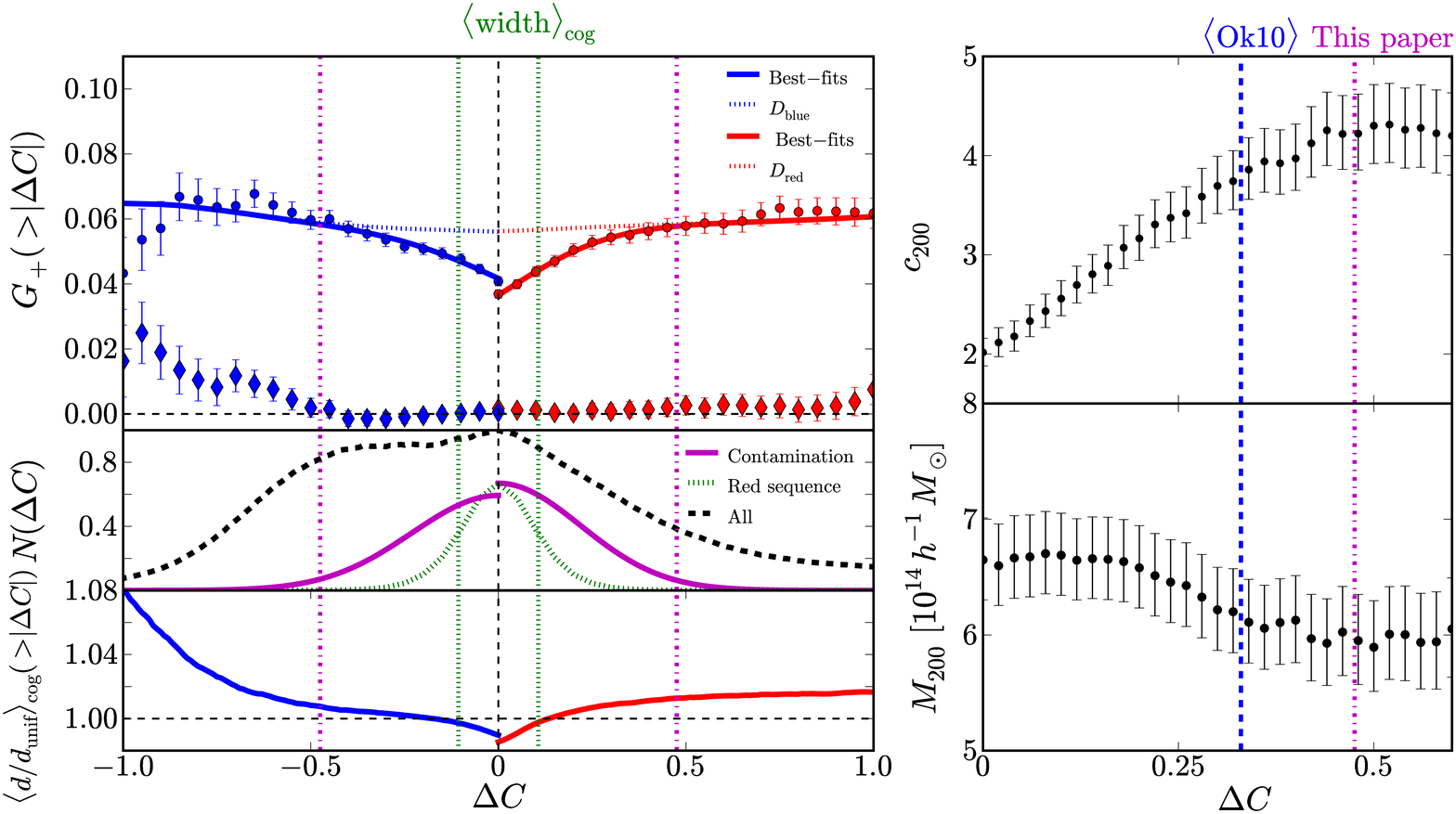}
  }
  \caption{{\sc Top-left}: Stacked reduced shear $G_+$ (Section 2) as a
    function of color offset from the stacked cluster red sequence
    (filled circles).  Filled diamonds show the mean shear calculated
    after rotating galaxies through $45^\circ$.  Solid curves show the
    best-fit lensing kernel plus contamination model described in Sec 2.
    The vertical dashed-dotted magenta line shows the color cut at
    which the fraction of contaminants is $1\%$.  {\sc Middle-left}:
    The color distribution of all galaxies at $21<i'<26$ (black dashed
    curve).  The width of the red sequence of bright ($i'<20$) cluster
    members is shown as the green dotted curve, with the green
    vertical lines de-marking the $1\sigma$ width of the red sequence.
    The magenta curves show the color distribution of contaminants in
    our model, and upon which the $1\%$ contamination cut is based.
    {\sc Bottom-left}: The mean distance with respect to the brightest
    cluster galaxy for 50 clusters, normalized by a uniform
    distribution. The faint blue population appears to be
    preferentially found at large cluster centric radii, suggesting
    that blue galaxy contamination may be dominated by galaxies in the
    cluster outskirts.  {\sc Right}: The run of $c_{200}$ (upper) and
    $M_{200}$ (lower) with $\Delta C$, showing the color cut that we
    adopt in this paper (magenta dot-dashed) and that of Ok10 (blue
    dashed).  Ok10 (see their Figure 14) chose their color cuts by eye
    based on the mean tangential distortion strength for the cluster
    sample.}
  \label{fig:dilution}
\end{figure*}

The mean tangential distortion strength averaged over (1) all galaxies
satisfying each color cut, (2) cluster-centric radii of
$0.1\hMpc<r<2.8\hMpc$, and (3) all 50 clusters,
$G_+\equiv\langle\langle\langle g_+\rangle\rangle\rangle$, increases
monotonically with $\Delta C$ for red galaxies
(Fig.~\ref{fig:dilution}).  We interpret the steep slope at
$0\simlt\Delta C\simlt0.3$ as arising from contamination by cluster
members.  Indeed, the mean cluster-centric radius of red galaxies is
an increasing function of $\Delta C$ at small $\Delta C$
(Fig.~\ref{fig:dilution}).  At $\Delta C\simgt0.4$ we interpret the
shallow slope of $G_+$ as arising from a slowly increasing redshift of
the faint red background population as $\Delta C$ increases.  We
therefore model the data with a Gaussian of width $\sigma$ centered at
$\Delta C=0$ (to represent the cluster population), and the mean
lensing kernel, $D(\Delta C)\equiv\langle D_{ls}/D_s\rangle$, for
galaxies in the COSMOS photometric redshift catalog
\citep{Ilbertetal:09} that matches each color cut.  The model for the
color-dependence of $G_+$ is therefore: $G_+(\Delta C)=A\,D(\Delta
C)\,(1-Bf(\Delta C))$, where $A$ converts $D$ into shear in a simple
manner, $B$ is the normalization of the Gaussian contaminant function
at $\Delta C=0$, and $f(\Delta C>0)=[1-{\rm erf}(\Delta
C/\sqrt{2}\sigma)]/2$.  This model has three free parameters: $A$,
$B$, and $\sigma$, and allows to estimate explicitly the fraction of
contaminant galaxies, $f$, as a function of $\Delta C$.

The best-fit model describes the red galaxies well
(Fig.~\ref{fig:dilution}, upper panel).  We conservatively adopt a
limit of $1\%$ on contaminating fraction, which translates into a
red color cut of $\Delta C>0.475$.  We select galaxies redder
than this cut for the results presented in Section 3; the mean number
density of these galaxies is $5.3\pm1.9{\rm arcmin}^{-2}$ per cluster,
where the uncertainty is the standard deviation among the 50 clusters.
We therefore achieve a total stacked number density of red galaxies of
$266.3{\rm arcmin}^{-2}$.

For completeness, we applied the same methods to blue galaxies,
describing the contaminating fraction as $f(\Delta C<0)=[1+{\rm
  erf}(\Delta C/\sqrt{2}\sigma)]/2$.  The model does not describe the
blue galaxies well, and we do not use them in Section \ref{sec:results}.

\section{Results}\label{sec:results}

Our results are based on stacking the \emph{red} background galaxy
sample, defined by $\Delta C>0.475$ (Section \ref{sec:analysis}), for
\emph{all} 50 clusters in the sample.

\subsection{Stacking and Modeling the Weak Shear Signal}\label{sec:stack}

\begin{figure*}
  \centerline{
    \includegraphics[width=0.8\textwidth]{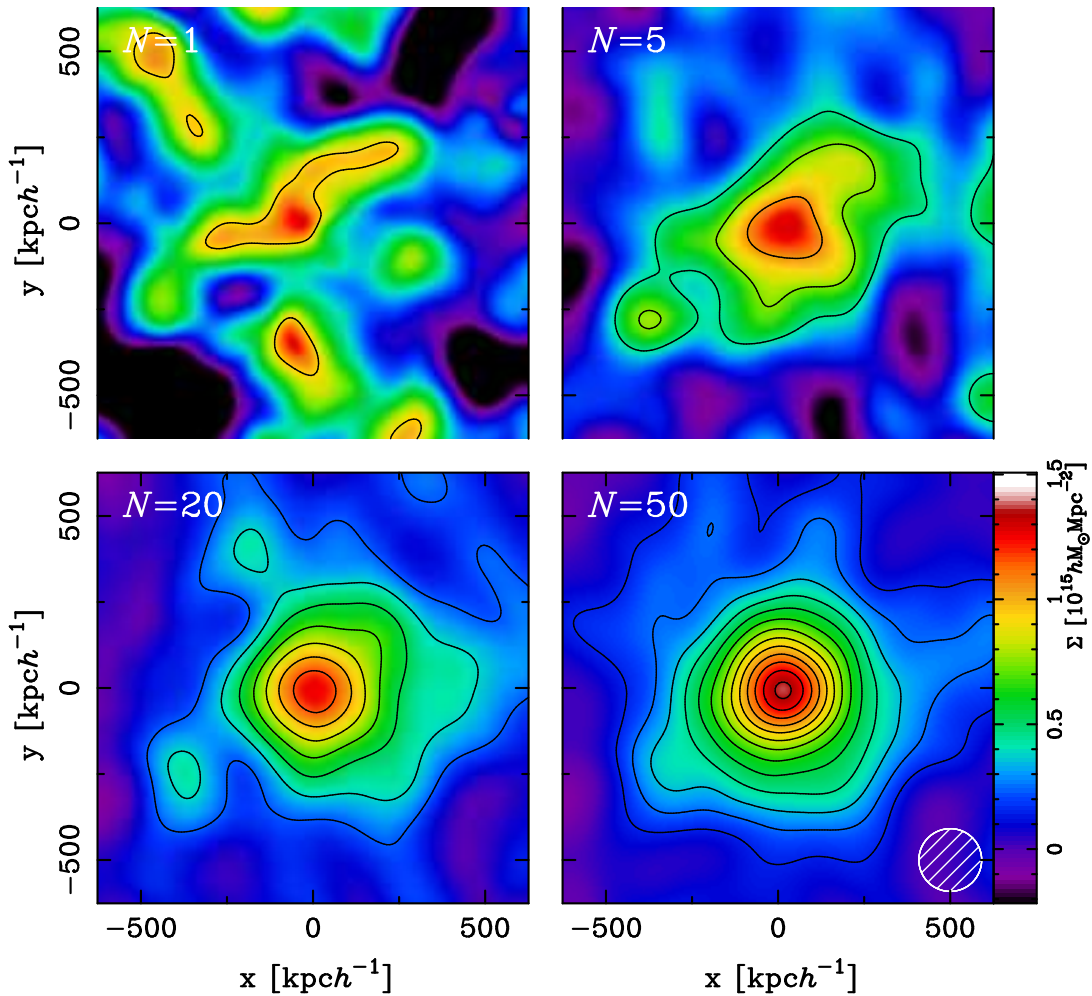}
  }
  \caption{ The projected mass distribution reconstructed from our
    weak-lensing catalogs, from one typical cluster ($N=1$;
    ABELL\,0141; upper left) to the full sample ($N=50$; bottom
    right).  Contours start at ${\rm S/N}=3$, and are spaced at
    $\Delta{\rm S/N}=2$.  A Gaussian smoothing scale of ${\rm
      FWHM}=2\,{\rm arcmin}$ is used in all panels (hatched region at
    lower right).  }
  \label{fig:massmap}
\end{figure*}

\begin{figure*}
  \centerline{
    \includegraphics[width=0.53\textwidth]{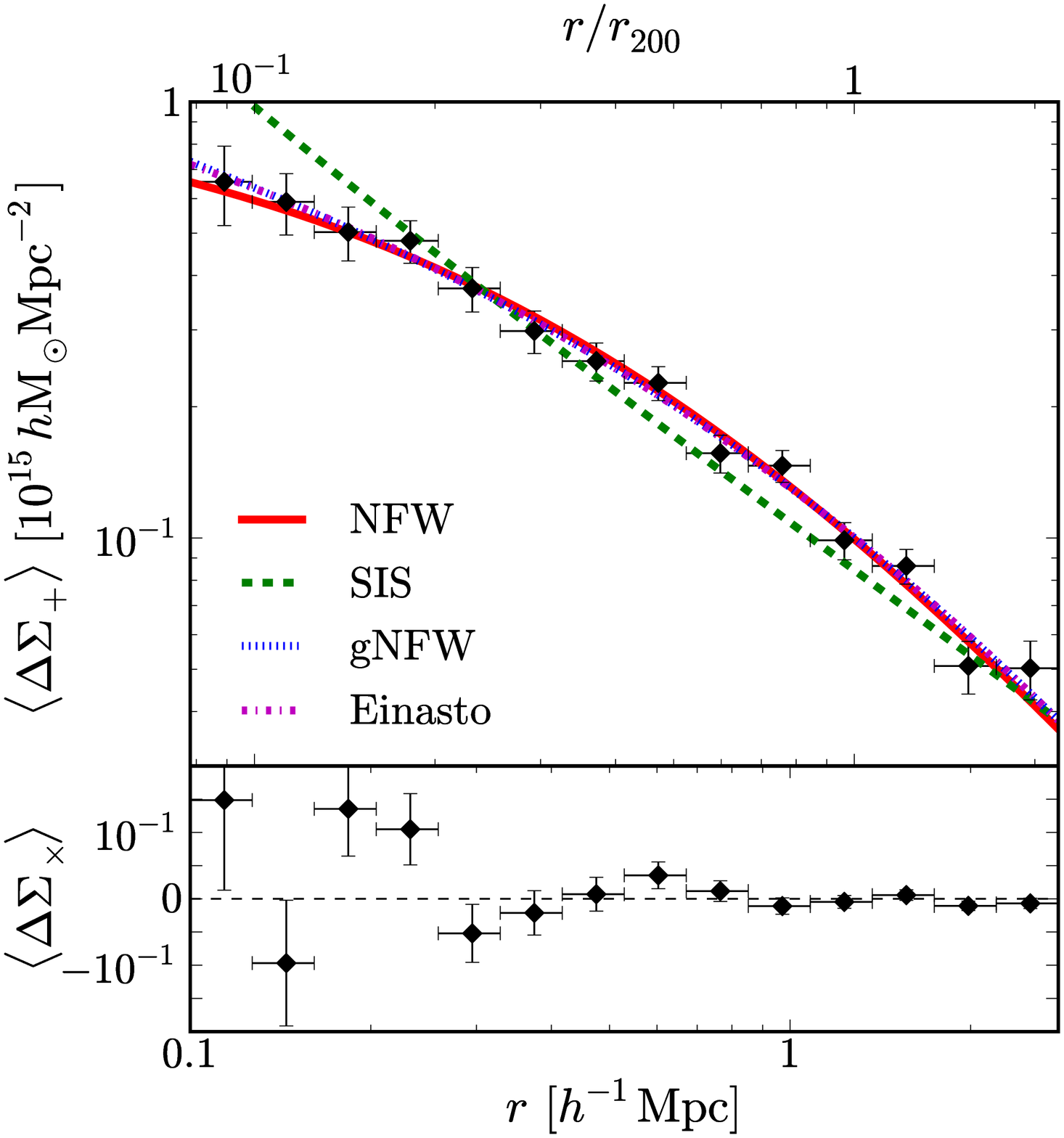}
    \includegraphics[width=0.5\textwidth]{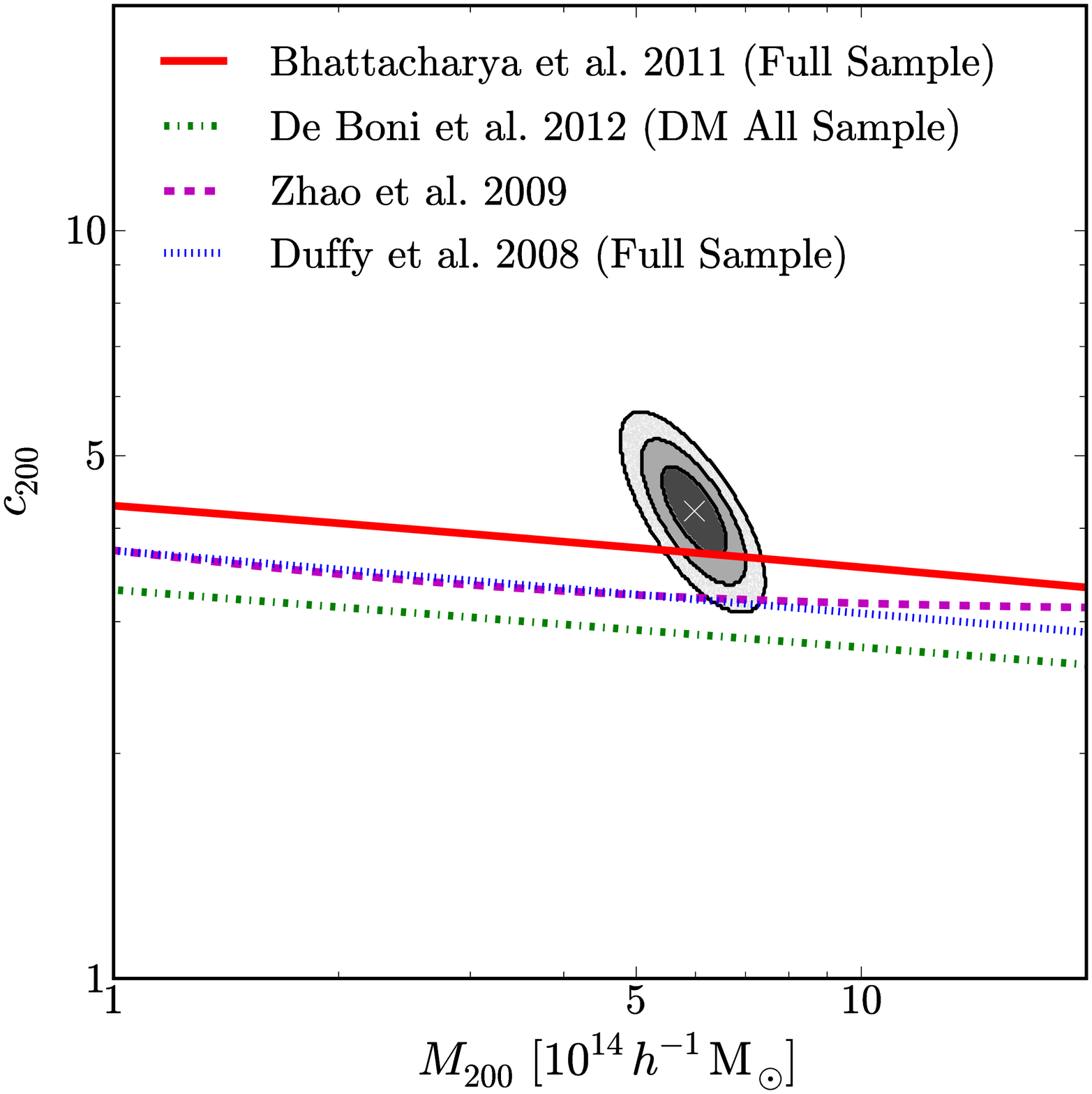} 
  }
  \caption{Stacked tangential shear profile of all 50 clusters in
    units of projected mass density, where different cluster and
    background galaxy redshifts galaxies are weighted by the lensing
    kernel \citep{Mandelbaumetal:06,Okabeetal:10,OguriTakada:11,
      Umetsuetal2011}.  The projected radius is computed from the
    weighted mean cluster redshift ($z_{\rm cluster}\simeq0.23$).  The
    solid, dashed, dotted and dashed-dotted curves are the best-fit
    Navarro-Frenk-White (NFW), singular isothermal (SIS), generalized
    NFW (gNFW) and Einasto profiles, respectively.  The lower panel
    shows the result of the $45^\circ$ test for systematic errors.
    {\sc Right} -- Stacked weak-lensing constraints on the mass and
    concentration of a complete volume-limited sample of 50 galaxy
    clusters at $\langle z\rangle=0.23$.  The white cross denotes the
    best-fit parameters and the contours show the 68.3\%, 95.4\%, and
    99.7\% confidence levels.  Note that the predicted relations have
    all been converted to be consistent with our analysis.}
  \label{fig:profile}
\end{figure*}

We detect each individual cluster at a typical peak signal-to-noise
ratio of ${\rm S/N}\simeq4$ in two-dimensional \cite{KaiserSquires:93}
mass reconstructions.  We also stack the shear catalogs in physical
length units centered on the respective BCGs and reconstruct the
average cluster mass distribution for the \emph{full} sample, with a
peak signal-to-noise ratio of ${\rm S/N}=28$ (Fig.~\ref{fig:massmap}).
Motivated by the symmetrical average mass map, we constructed the
stacked tangential shear profile for the full sample
(Fig.~\ref{fig:profile}) following the procedure of
\citet{Umetsuetal2011}.  In brief, we center the catalogs on the
respective BCGs, and stack in physical length units across the radial
range $100\hkpc<r<2.8\hMpc$, in 14 log-spaced bins.  We detect the
signal at ${\rm S/N}=32.7$, using the full covariance matrix to take
into account projected uncorrelated large-scale structure and
intrinsic ellipticity noise \citep[e.g.][]{Hoekstra2003,
  Hoekstraetal2011, OguriTakada:11, Umetsuetal2011, Oguri12},
computing the cosmic-shear contribution using the non-linear matter
power spectrum \citep{Smithetal2003} for the WMAP7 cosmology and the
shape noise from the diagonal matrix.  The $45^\circ$-rotated
distortion component is consistent with a null signal, confirming that
residual systematic errors are at least an order of magnitude smaller
than the measured lensing signal.

\begin{deluxetable*}{l|lccccc}
  \tablewidth{0pt}
  \tabletypesize{\small}
  \tablecaption{Density Profile Models \label{tab:mass}}
  \startdata \hline\hline 
  Model
  & Shape parameter\tablenotemark{a} 
  & $M_{\rm vir}$
  & $c_{\rm vir}$\tablenotemark{b} 
  & $M_{200}$
  & $c_{200}$\tablenotemark{b} 
  & $\chi^2_{\rm min}/{\rm d.o.f}$
  \\
  & 
  & ($10^{14}h^{-1}M_\odot$)
  &
  & ($10^{14}h^{-1}M_\odot$)
  & \\
  \hline
  NFW   
  & $\gamma=1$ 
  & $7.19_{-0.50}^{+0.53}$
  & $5.41_{-0.45}^{+0.49}$ 
  & $5.98_{-0.38}^{+0.40}$
  & $4.22_{-0.36}^{+0.40}$ 
  & $7.2/12$\\
  gNFW  & $\gamma=1.27_{-0.37}^{+0.24}$ 
  & $7.50_{-0.65}^{+0.74}$
  & $4.88_{-0.86}^{+0.86}$
  & $6.15_{-0.44}^{+0.48}$
  & $3.79_{-0.69}^{+0.69}$
  & $6.6/11$ \\
  Einasto & $\alpha=0.188_{-0.058}^{+0.062}$
  & $7.49_{-0.73}^{+0.86}$ 
  & $4.92_{-0.80}^{+0.57}$
  & $6.15_{-0.45}^{+0.50}$
  & $3.82_{-0.66}^{+0.48}$
  & $6.6/11$ 
  \enddata
  \tablenotetext{a}{Parameter describing the shape of the mass density
    profile on small scales.}  
  
  \tablenotetext{b}{NFW-like concentration parameter defined by
    $c_\Delta^{\rm NFW}=r_{\Delta}/r_{\rm s}$, $c_{\rm -2}^{\rm
      gNFW}=(r_{\Delta}/r_{\rm s})/(2-\gamma)$ and $c_{\rm \Delta}^{\rm
      Einasto}=r_{\Delta}/r_{-2}$.}
  
\end{deluxetable*}

The stacked shear profile (Fig~\ref{fig:profile}) is well-described by
the so-called NFW profile: $\rho\propto x^{-1}(1+x)^{-2}$, where
$x\equiv r/r_{\rm s}$, and $d\log\rho/d\log r=-2$ at $r=r_{\rm s}$
\citep[][]{NFW97}.  We express our model fits in terms of the virial
mass $\Mvir\equiv(4\pi/3)\rho_{\rm cr}\Dvir\rvir^3$, and the
concentration parameter $\cvir\equiv r_{\rm vir}/r_{\rm s}$, where
$\Dvir$ is the virial over-density and $\rho_{\rm cr}$ is the critical
density.  We measure both parameters to sub-10\% statistical precision
(Table~\ref{tab:mass}), obtaining a best fit concentration parameter
of $\cvir=5.4\pm0.5$.  Indeed, the statistical errors on concentration
are comparable with the differences between the predictions from
different numerical simulations (Fig.~\ref{fig:profile}).  Moreover
the observed concentration parameter exceeds the predicted
concentration from numerical simulations \citep{Duffyetal:08,
  Zhaoetal2009, Bhattacharyaetal2011, DeBonietal2012}.

The NFW model fit described above does not motivate fitting more
flexible models to our data (Table~\ref{tab:mass}).  Nevertheless, for
completeness, we fit the generalized NFW (gNFW) and \cite{Einasto1965}
profiles.  The former adds a free parameter $\gamma$ to the NFW
profile: $\rho\propto{x}^{-\gamma}(1+x)^{-3+\gamma}$; the latter
describes the shape of the profile slope thus: $d\log\rho/d\log
{r}=-2(r/r_{-2})^{\alpha}$.  The best-fit gNFW profile is consistent
with NFW, with $\gamma=1.27^{+0.24}_{-0.37}$.  The best-fit Einasto
profile has $\alpha=0.19\pm0.06$, consistent with numerical
simulations, e.g.\ $\langle\alpha\rangle=0.175\pm0.046$
\citep{Gaoetal2012}, and $\langle\alpha\rangle=0.183$
\citep{Navarroetal2004}.  We also measure the inner slope of the best
fit density profile models directly, obtaining
$\beta(r=0.01r_{200})=-d\log\rho/d\log{r}=1.1$ for the gNFW and
Einasto models, in good agreement with $\langle\beta\rangle\simeq1.1$
\citep{Navarroetal2004,Gaoetal2012}.

We also examine the possible impact of adiabatic contraction on the
total measured density profile \citep[e.g.][]{Gnedinetal:04} by
introducing a central point mass into the model.  We obtain an upper
limit on the point mass of $M_{\rm point}\simlt12\times10^{12}\hMsol$,
which is degenerate with the structural parameters of the smooth
component in all models (NFW, gNFW, and Einasto).  The best-fit mass
and concentration parameters do not change significantly from those
listed in Table~1.  The excellent fit of the NFW model -- that is
based on numerical dark matter only simulations -- to our weak-lensing
data, and the results of adding baryons to the model (albeit in a
simplified form) \emph{suggest} the dark matter may not suffer
adiabatic contraction by baryons in the cluster core.  We will return
to this topic in a future article that combines strong- and
weak-lensing constraints.

\subsection{Systematic Errors}\label{sec:syst}

We investigate the sensitivity of our results to systematic errors.
In summary, we conclude that systematic errors are sub-dominant to the
statistical errors discussed in Section\ref{sec:stack}.  

\smallskip

\noindent
\emph{Shear calibration} -- We confirmed the reliability of our shape
measurements using simulated data that were generated using {\sc
  glafic} \citep{Oguri10a} with point spread functions described by
the Moffat profile with a range of seeing ($0\farcs5<{\rm
  FWHM}<1\farcs1$) and power indices ($3<\beta<12$), as described in
\cite{Oguri12}.  We obtain a multiplicative calibration bias ($m$) and
additive residual shear offset ($c$) \citep[defined
following][]{Heymans06} of $|m|\simlt0.03$ and
$|c|\simlt2\times10^{-4}$, respectively, for ${\rm
  FWHM}\simeq0\farcs7$.

\smallskip

\noindent
\emph{Radial and color cuts} -- Our results change by just $\Delta
c_{\rm vir}\simeq0.1$ when we vary the number of bins between 8 and
18, change the inner radial cut from 80 to $200\hkpc$ or the outer
radial cut between 2.5 and $3.5\hMpc$.  The stability of our results
under variations of the inner radial cut underlines the robustness of
our new approach to selecting red galaxies, and the negligible level
of $\langle\Sigma_\times\rangle$ noted in Section \ref{sec:stack}.
Moreover, the constraints on concentration are stable to $\Delta
c_{\rm vir}\simlt0.2$ with respect to increasing the color cut beyond
$\Delta C>0.475$, and to fitting only to galaxies brighter than
$i'=25$.  The constraints on $\Mvir$ are stable to a few per cent
under the same tests (Fig.~\ref{fig:dilution}).

\smallskip

\noindent
\emph{Stacking procedure: radial bins} -- We construct synthetic weak
shear catalogs based on analytic NFW halos that match the
mass-concentration relation predicted from numerical simulations.
These catalogs match the observed number density and field of view of
our Subaru data.  We draw 300 samples of 50 clusters from the
predicted cluster distribution, and stack the respective shear
profiles in both physical length units (as in Section \ref{sec:stack}) and
length units scaled to $r_{200}$ of each halo.  We do not detect any
bias in the measured mean concentration of the stacked clusters,
obtaining $\langle c/c_{\rm truth}\rangle=1.02\pm0.07$ for stacking in
physical length units, and find $\langle c/c_{\rm
  truth}\rangle=1.08\pm0.07$ for re-scaled length units.  In both
cases we obtain $\langle M/M_{\rm truth}\rangle=0.96\pm0.06$; the
uncertainties are the standard deviation on the 300 samples of 50
  clusters.  The non-detection of a systematic error arising from
stacking in physical units is consistent with Ok10's result that their
mass-concentration relations from individual and stacked clusters
(using physical length units) are self-consistent.  We also note that
stacking in re-scaled length units weights the contribution of each
cluster to each bin in a nonlinear and model-dependent manner:
$w\propto\theta\Delta\theta\propto r_{200}^2\propto M_{200}^{2/3}$.

Real clusters are aspherical, embedded in the large-scale-structure,
and contain baryons.  As numerical hydrodynamical simulations become
more realistic, robust tests based on simulated clusters should
therefore become possible.  We conduct a preliminary test using
clusters extracted from the new ``Cosmo-OWLS'' simulation, that
implements the AGN model described in \citet{McCarthyetal:11} in a
$400\hMpc$ box, with weak-lensing catalogs constructed following
\citet{2012MNRAS.421.1073B}.  The results are consistent with the
analytic NFW tests -- i.e.\ we do not detect any systematic error on
the measurement of concentration based on stacking in physical length
units.

\smallskip

\noindent
\emph{Stacking procedure: centering} -- We also checked whether the
results are affected by adopting the BCG as the center of each
cluster, by adding an off-centering parameter $\sigma_{R_{\rm off}}$
to the models following \citet{Johnstonetal:07}.  The best-fit $M_{\rm
  vir}$ and $c_{\rm vir}$ are unchanged, and we obtain an upper limit
of $\sigma_{R_{\rm off}}<29\hkpc$.

\subsection{Comparison with Okabe et al.\ (2010)}

We fit an NFW model to Ok10's stacked red$+$blue catalog and our own
stacked red galaxy catalog for the 21 clusters in common between the
two studies, finding that our mean masses and concentrations are
  $\sim14-20\%$ and $\sim15-17\%$ greater than theirs
  (Table~\ref{tab:comp}).  The main differences between Ok10 and our
  analysis relate to color-selection of background galaxies, and their
  shape measurement methods (\S\ref{sec:analysis}).  We attribute the
  differences between our respective mass measurements mainly to a
  combination of (1) contamination of Ok10's blue galaxy sample at
  large cluster-centric radii and (2) systematics in Ok10's shape
  measurement methods.  We attribute the differences between the
  respective concentration measurements mainly to contamination of
  Ok10's red galaxy catalog -- their less conservative red color cut
  ($\langle\Delta C\rangle=0.33$) leads to an overall $\sim5\%$
  contamination by galaxies that preferentially lie at small
  cluster-centric radii (see right panel of Fig.~1).  We note that the
  results in this section are consistent with
  \citet{2013A&A...550A.129P} and \citet{2012arXiv1208.0605A}.

\begin{deluxetable}{l|ccc}
  \tablewidth{0pt}
  \tabletypesize{\small}
  \tablecaption{Comparison with Okabe et al. (2010)\label{tab:comp}}
  \startdata \hline\hline 
   \noalign{\smallskip}
 & \multispan3{\hfill Over-density, $\Delta$\hfill} \\
Parameters\tablenotemark{a}   & $\Dvir$
  & $200$
  & $500$ \\  \hline
$M_{\Delta}^{2013}/M_{\Delta}^{2010}$  & $1.14\pm0.16$
  & $1.16\pm0.14$
  & $1.20\pm0.12$ \\
$c_{\Delta}^{2013}/c_{\Delta}^{2010}$  & $1.15\pm0.19$
  & $1.16\pm0.19$ 
  & $1.17\pm0.22$ 
  \enddata
  \tablenotetext{a}{Ratio of the stacked mass and concentration
      obtained from our methods and those of Ok10, for the 21 clusters
      in common between the two studies.}
\end{deluxetable}

\section{Summary}\label{sec:summary}

We have used sensitive high resolution observations with Subaru to
measure the average density profile of an approximately mass-selected
sample of 50 galaxy clusters at $0.15<z<0.3$.  Careful treatment of
systematic errors indicates that they are all smaller than the
statistical errors.  In particular, we achieve just 1\% contamination
of the background galaxy sample by foreground and cluster galaxies,
tests on simulated data indicate that our shape measurement
multiplicative systematic error is $m\simlt0.03$, and errors from
choice of binning scheme are just a few per cent.  When the signal
from all 50 clusters is combined together we achieve a number density
of background galaxies of $266.3{\rm arcmin}^{-2}$.

The shape of the stacked density profile is consistent with numerical
simulations across the radial range $100\hkpc-2.8\hMpc$.
Specifically, we find no statistical evidence for departures from the
NFW profile.  We constrain the mean mass and concentration of the
clusters to sub-10\% precision, obtaining $c_{\rm
  vir}=5.41^{+0.49}_{-0.45}$.  This level of precision is comparable
with the differences between the concentrations predicted by different
numerical simulations, and therefore opens the possibility of
discriminating between different simulations using observational data
in the near future.  

Our results emphasize the power of stacked weak-lensing for
constraining the average mass and shape of galaxy clusters.  Surveys
including Hyper Suprime-Cam on Subaru, the Dark Energy Survey, and
KIDS, all hold much promise for stacked weak-lensing studies of less
massive clusters, including those at higher redshifts.  However
significant advances on the precision that we have achieved here on
massive low redshift clusters await future facilities such as LSST and
\emph{Euclid} to provide the required number density of background
galaxies on these rare and massive low redshift clusters.

\section*{Acknowledgments}

We thank Ian McCarthy, Yannick Bah\'e, and Joop Schaye for sharing
their weak shear catalogs from the Cosmo-OWLS simulation in advance of
publication.  We also thank our LoCuSS colleagues, especially Dan
Marrone, Gus Evrard, Pasquale Mazzotta, Arif Babul, and Alexis
Finoguenov for many helpful discussions and comments.  We acknowledge
the Subaru Support Astronomers, plus Paul May, Chris Haines, and
Mathilde Jauzac, for assistance with the Subaru observations.  We are
grateful to N.\ Kaiser and M.\ Oguri for making their {\sc imcat} and
{\sc glafic} packages public.  This work is supported in part by
Grant-in-Aid for Scientific Research on Priority Area No.\ 467
``Probing the Dark Energy through an Extremely Wide \& Deep Survey
with Subaru Telescope'', by World Premier International Research
Center Initiative (WPI Initiative), MEXT, Japan, and by the FIRST
program ``Subaru Measurements of Images and Redshifts (SuMIRe)''.  GPS
acknowledges support from the Royal Society.  KU acknowledges partial
support from the National Science Council of Taiwan (grant
NSC100-2112-M-001-008-MY3) and from the Academia Sinica Career
Development Award.



\end{document}